\documentclass[lettersize,journal]{IEEEtran}
\usepackage{amsmath,amsfonts}
\usepackage{algorithmic}
\usepackage{algorithm}
\usepackage{array}
\usepackage[caption=false,font=normalsize,labelfont=sf,textfont=sf]{subfig}
\usepackage{textcomp}
\usepackage{stfloats}
\usepackage{url}
\usepackage{verbatim}
\usepackage{graphicx}
\usepackage{cite}
\usepackage{color}
\usepackage{booktabs}
\usepackage{makecell}
\usepackage{bm}
\usepackage{csquotes}

\begin{document}
	
\title{Learning-based Block-wise Planar Channel Estimation for Time-Varying MIMO OFDM}

\author{Chenchen Liu, Wenjun Jiang,~\IEEEmembership{Graduate Student Member,~IEEE}, and Xiaojun Yuan,~\IEEEmembership{Senior Member,~IEEE}
}
\maketitle

\begin{abstract}
In this paper, we propose a learning-based block-wise planar channel estimator (LBPCE) with high accuracy and low complexity to estimate the time-varying frequency-selective channel of a multiple-input multiple-output (MIMO) orthogonal frequency-division multiplexing (OFDM) system. First, we establish a block-wise planar channel model (BPCM) to characterize the correlation of the channel across subcarriers and OFDM symbols. Specifically, adjacent subcarriers and OFDM symbols are divided into several sub-blocks, and an affine function (i.e., a plane) with only three variables (namely, mean, time-domain slope, and frequency-domain slope) is used to approximate the channel in each sub-block, which significantly reduces the number of variables to be determined in channel estimation. Second, we design a 3D dilated residual convolutional network (3D-DRCN) that leverages the time-frequency-space-domain correlations of the channel to further improve the channel estimates of each user. Numerical results demonstrate that the proposed significantly outperforms the state-of-the-art estimators and maintains a relatively low computational complexity.  	
\end{abstract}

\begin{IEEEkeywords}
MIMO-OFDM, channel estimation, low-complexity, time-varying frequency-selective channels 	
\end{IEEEkeywords}

\section{Introduction}
Multiple-input multiple-output (MIMO) orthogonal frequency-division multiplexing (OFDM) is a key enabling technology for fifth-generation (5G) and beyond-5G (B5G) communication systems\cite{wang2023road}. A fundamental problem in a MIMO-OFDM system is channel estimation\cite{wang2023road,chen20235g}. With the increase in the numbers of access users, MIMO antennas, and OFDM subcarriers, the computational complexity of channel estimation significantly increases and becomes a heavy burden in system implementation\cite{chataut2020massive,damsgaard2023approximate}. As such, considering the time-varying and frequency-selective characteristics of the channel, it is an outstanding challenge to design a high-accuracy and low-complexity channel estimator for a practical MIMO-OFDM system. 

Recently, deep learning has shown its great potential in channel estimation\cite{ma2020data,mashhadi2021pruning,dong2019deep,liao2019deep,soltani2019deep,melgar2022deep,sun2021icinet,9837303,he2018deep,LTMP}. The learning-based low-complexity channel estimations usually consist of two steps \cite{liao2019deep,soltani2019deep,melgar2022deep,sun2021icinet,9837303,he2018deep,LTMP}. In the first step, the received signal is divided into multiple sub-blocks, and the channels per sub-block are estimated, e.g., by the least square (LS) method \cite{liao2019deep,soltani2019deep,melgar2022deep,sun2021icinet}. This block-wise scheme reduces the size of matrix/vector multiplication for complexity reduction and meanwhile separates channels of different users per sub-block. Then, in the second step, the channels on all sub-blocks are combined and refined, e.g., by using neural networks to exploit the channel correlations\cite{liao2019deep,soltani2019deep,melgar2022deep,sun2021icinet,9837303,he2018deep,LTMP}. These existing approaches, however, have their own limitations. For example, in the first step, the flat-fading assumption imposed on each sub-block of channels \cite{liao2019deep,soltani2019deep,melgar2022deep,sun2021icinet} may lead to serious model mismatch, especially in fast-fading and highly frequency-selective channels. In the second step, the existing approaches\cite{9837303,liao2019deep,soltani2019deep,melgar2022deep,sun2021icinet,he2018deep,LTMP} typically exploit channel correlations in the time domain, frequency domain, and/or space domain, but not all of them together. How to efficiently exploit the three-domain correlations simultaneously with an affordable complexity remains an outstanding challenge.

To improve the channel estimation performance and maintain low complexity, we propose a learning-based block-wise planar channel estimator (LBPCE) for time-varying MIMO-OFDM systems. As a low-complexity channel estimator, LBPCE also consists of two steps. In the first step, we develop a block-wise planar channel model (BPCM) to approximate the time-varying frequency-selective channel in each sub-block with three model coefficients. Based on the BPCM, we employ the linear minimum mean-squared error (LMMSE) principle to estimate the model coefficients, and then reconstruct the channel estimates. In the second step, we design a 3D dilated residual convolutional network (3D-DRCN) to filter the channel estimates of each user, which exploits the correlations of the channel in the time-frequency-space domain. Simulation results show that the proposed algorithm significantly outperforms the state-of-the-art approaches with a comparable computational complexity.

\section{System Model}
\subsection{Multiuser MIMO-OFDM Channel}
\begin{figure}[h]
	\centering
	\includegraphics[width=3.2in]{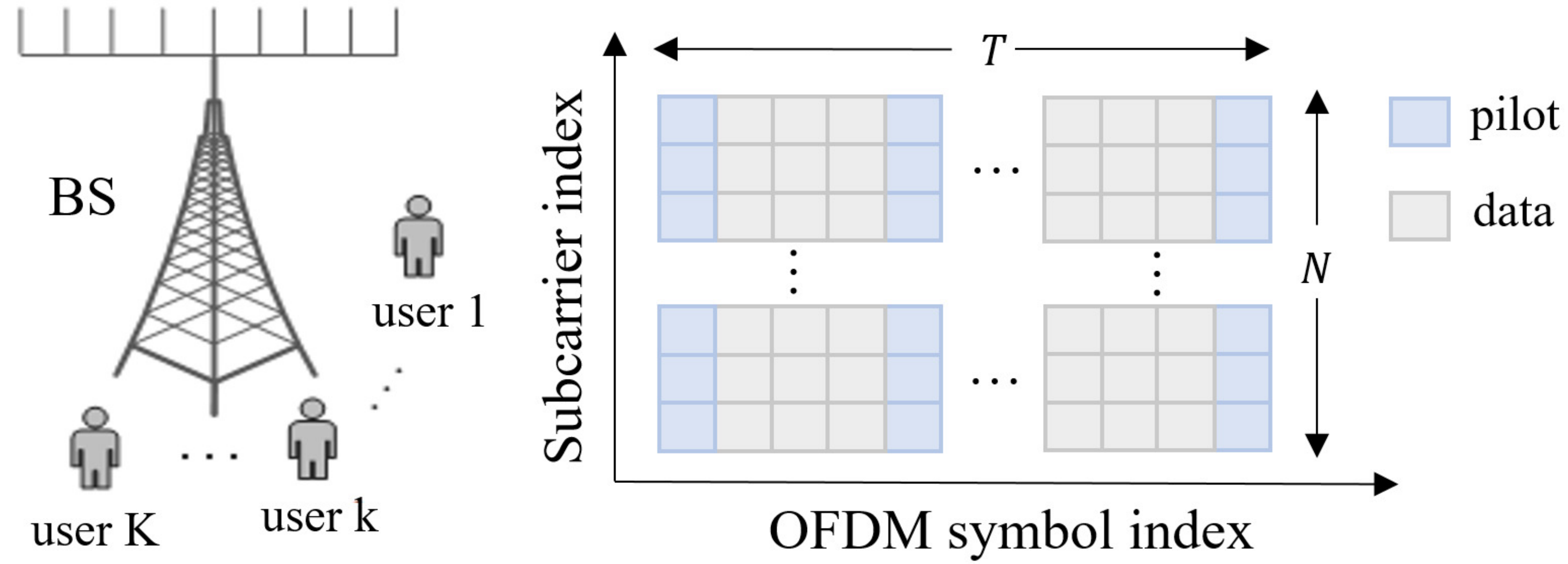}
	\caption{An illustration of the multiuser MIMO-OFDM system.}
	\label{subframe}
\end{figure}

Consider an uplink multiuser MIMO-OFDM system as illustrated in Fig. \ref{subframe}, where an $M$-antenna base station (BS) serves $K$ single-antenna users. Each OFDM frame comprises $N$ subcarriers and $T$ OFDM symbols. Within a frame, $T_{\mathrm P}$ symbols are used for pilot transmission and the remaining $T-T_{\mathrm P}$ symbols are used for data transmission. 

We next establish a time-frequency-domain system model. In each frame, we divide the $T$ OFDM symbols into $U$ sub-blocks and the $N$ subcarriers into $V$ sub-blocks. In the $(u,v)$-th sub-block, the channel coefficient of the $k$-th user at the $m$-th antenna on the $t$-th OFDM symbol and the $n$-th subcarrier is given by \cite{bajwa2010compressed}  \cite{3GPP_TS_38_901_v17.0.0}
\begin{gather} 
	 h_{u,v,k,m,t,n} = \sum_{l=1}^{L_k} \beta_{k,m,l}  e^{-j2 \pi \left( \Delta f \tau_{k,l} ((v-1)\frac{N}{V}+n) + v_{k,l}t \right)},  \notag  \\
	 k = 1,...,K;  m = 1,...,M; t = 1,...,\frac{T}{U}; n = 1,...,\frac{N}{V}, 
\end{gather}
where $L_k$ is the path number of the $k$-th user; $\beta_{k,m,l}$ is the channel coefficient of the $k$-th user at the $m$-th antenna on the $l$-th path; $\Delta f$ is the subcarrier spacing; $\tau_{k,l}$ and $v_{k,l}$ are the delay and the Doppler shift  of the $k$-th user on the $l$-th path, respectively. In the $t$-th symbol, denote by $\mathbf x_{u,v,k,t} = [x_{u,v,k,t,1},...,x_{u,v,k,t,N/V}]^T \in \mathbb C^{\frac{N}{V} \times 1}$ the transmitted signal vector of the $k$-th user with $x_{u,v,k,t,n}$ being the transmitted signal of the $k$-th user on the $n$-th subcarrier; denote by $\mathbf h_{u,v,k,m,t} = [h_{u,v,k,m,t,1},...,h_{u,v,k,m,t,N/V} ]^T \in \mathbb C^{\frac{N}{V} \times 1}$ the channel vector of the $k$-th user at the $m$-th antenna with $h_{u,v,k,m,t,n}$ being the channel of the $k$-th user at the $m$-th antenna on the $n$-th subcarrier. We assume that the length of the cyclic prefix (CP) exceeds the maximum channel delay. After removing the CP and applying the discrete Fourier transform to the received signal, the signal model in the $t$-th OFDM symbol of the $(u,v)$-th sub-block at the $m$-th receive antenna is expressed as
\begin{align}
	\mathbf y_{u,v,m,t} & = \sum_{k=1}^{K} {\rm diag} (\mathbf x_{u,v,k,t}) \mathbf h_{u,v,k,m,t} +\mathbf w_{u,v,m,t}, 
	\label{SystemModelMatrix_symbol}
\end{align}
where $\rm diag(\cdot)$ represents the diagonalization operation; $\mathbf y_{u,v,m,t} \in \mathbb C^{\frac{N}{V} \times 1}$ is the received signal at the $m$-th antenna and the $t$-th symbol; $\mathbf w_{u,v,m,t} \in \mathbb C^{\frac{N}{V} \times 1}$ is the additive white Gaussian noise (AWGN) with elements independently drawn from $\mathcal{CN}(0,\sigma^2)$.\footnote{We assume that the channel variation within the time duration of an OFDM symbol is very small, so the inter-carrier interference is ignored.}

For the $(u,v)$-th sub-block, we further donote $\mathbf x_{u,v,k} = [\mathbf x_{u,v,k,1}^T,..., \mathbf x_{u,v,k,T/U}^T]^T \in \mathbb C^{\frac{NT}{VU} \times 1}$ and $\mathbf H_{u,v,k} = [\mathbf h_{u,v,k,1},...,\mathbf h_{u,v,k,M}] \in \mathbb C^{\frac{NT}{VU} \times M}$ with $\mathbf h_{u,v,k,m} = [\mathbf h_{u,v,k,m,1}^T,...,\mathbf h_{u,v,k,m,T/U}^T]^T$. We obtain the signal model as
\begin{align}
	\mathbf Y_{u,v} & = \sum_{k=1}^{K} {\rm {diag}} (\mathbf x_{u,v,k})   \mathbf H_{u,v,k} +\mathbf W_{u,v}, 
	\label{SystemModelMatrix}
\end{align}
where $\mathbf Y_{u,v} = [\mathbf y_{u,v,1},...,\mathbf y_{u,v,M}] \in \mathbb C^{\frac{NT}{VU} \times M}$ is the received signal of $M$ antennas with $\mathbf y_{u,v,m} = [\mathbf y_{u,v,m,1}^T,...,\mathbf y_{u,v,m,T/U}^T]^T$; $\mathbf W_{u,v} \in \mathbb C^{\frac{NT}{VU} \times M}$ is the AWGN.

From \eqref{SystemModelMatrix}, we can extract rows of $\mathbf x_{u,v,k}$, $\mathbf H_{u,v,k}$, and $\mathbf Y_{u,v}$ to obtain the system model for the pilot symbols. To this end, denote by  $\mathbf S_{u,v} \in \mathbb C^{ \frac{NT_{\mathrm P}}{VU} \times \frac{NT}{VU}}$ the row selection matrix of the $(u,v)$-th sub-block. Define the set of the pilot positions in the $(u,v)$-th sub-block is $ \mathcal T_{{\mathrm P} u,v} = \{i_t|t=1,...,T_{\mathrm P}/U, i_t=1,...,T/U\}$. The $((t-1)\frac{N}{V}+n,(i_t-1)\frac{N}{V}+n)$-th entry of $\mathbf S_{u,v}$ is 1 and 0 otherwise, where $t=1,...,T_{\mathrm P}/U$, $i_t \in \mathcal T_{{\mathrm P} u,v}$, $n=1,...,N/V$. Then, the transmitted pilot signal is given by $\mathbf x_{u,v,k}^{\mathrm P} = \mathbf S_{u,v} \mathbf x_{u,v,k} \in \mathbb C^{\frac{NT_{\mathrm P}}{VU} \times 1}$. The pilot channel is given by $\mathbf H_{u,v,k}^{\mathrm P} = \mathbf S_{u,v} \mathbf H_{u,v,k} \in \mathbb C^{\frac{NT_{\mathrm P}}{VU} \times M}$. And the received pilot signal is given by $\mathbf Y_{u,v}^{\mathrm P} = \mathbf S_{u,v} \mathbf Y_{u,v} \in \mathbb C^{\frac{NT_{\mathrm P}}{VU} \times M}$.

\subsection{Existing Approaches}
We aim to estimate the channel $\{ \mathbf H_{u,v,k} \}$ from the pilot measurements $\{\mathbf Y_{u,v}^{\mathrm P}\}$. A Bayesian optimal method is the minimum mean-squared error (MMSE) estimator which exploits the time-frequency-space-domain channel correlations. However, the computational complexity of MMSE estimation is $\mathcal{O}(T_{\mathrm P}^3N^3M^3)$, which is highly burdensome for a relatively large $T_{\mathrm P}$, $N$, and $M$.

\begin{figure}[h]
	\centering
	\includegraphics[width=3.1in]{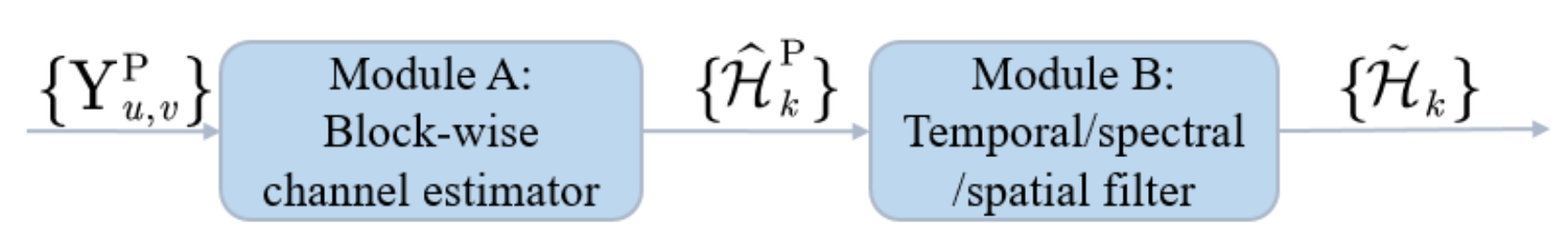}
	\caption{The block diagram of the existing low-complexity approaches.}
	\label{lowcomplexity}
\end{figure}

To reduce complexity, a popular approach is to divide the overall channel estimator into two concatenated modules as shown in Fig. \ref{lowcomplexity}. In module A, the received pilot signals are divided into UV block-wise pilot measurements $ \{\mathbf Y^{\mathrm P}_{u,v} \}$ to estimate $\{ {\mathbf H}_{u,v,k}^{\mathrm P}\}^{K}_{k=1}$ in a block-wise manner, i.e., the estimate of ${\mathbf H}_{u,v,k}^{\mathrm P}$, denoted by $\hat {\mathbf H}_{u,v,k}^{\mathrm P}$, for each user $k$ is obtained solely based on the observed $\mathbf Y_{u,v}^{\mathrm P}$. Then, the channel estimates $\{\hat{\mathbf H}_{u,v,k}^{\mathrm P}\}_{u=1,v=1}^{U,V}$ are combined as the coarse estimate $\hat{\boldsymbol{\mathcal H}}^{\mathrm P}_k \in \mathbb C^{ T_{\mathrm P} \times N \times M}$ for each user $k$. In module B, each $\hat{\boldsymbol{\mathcal H}}^{\mathrm P}_k$ passes through a filter to obtain a refined channel estimate $\tilde{\boldsymbol{\mathcal H}}_k \in \mathbb C^{ T \times N \times M}$ by exploiting the time/frequency/space correlations of the channel of the user $k$.

The existing approaches \cite{liao2019deep,soltani2019deep,melgar2022deep,sun2021icinet,9837303,he2018deep,LTMP}, however, have their own limitations. Specifically, in module A, oversimplified assumptions are imposed on channel ${\mathbf H}_{k,u,v}^{\mathrm P}$ for convenience of computation. For example, the authors in \cite{liao2019deep,soltani2019deep,melgar2022deep,sun2021icinet}, use the LS method by assuming that the channel within each sub-block is flat fading, which introduces a substantial model mismatch,  especially in a fast-varying rich-scattering environment. In module B, the channel correlations in the time-frequency-space-domain are not fully exploited. For example, the authors in \cite{9837303} ignore the space-domain correlation for the multiantenna system, and the authors in \cite{he2018deep} \cite{LTMP} ignore the time-domain correlation for the time-varying channels. To address the above issues and further improve the performance, we propose LBPCE as detailed in the next section. 

\section{Learning-Based Block-wise Planar Channel Estimation}
In this section, we describe LBPCE in detail. First, for module A, to reduce the model mismatch introduced by the flat-fading assumption, we develop a BPCM with only three variables in each sub-block to more accurately approximate the time-varying frequency-selective channel. Based on BPCM, we use the LMMSE principle to acquire the model variables in each sub-block. Subsequently, to fully exploit the correlations of the channel in the time-frequency-space-domain, we design a low-complexity 3D-DRCN as the filter of module B.

\subsection{Channel Estimation with BPCM}
In a time-varying frequency-selective channel, the variations of the channel coefficients across the subcarriers and OFDM symbols are continuous. Inspired by \cite{jiang2022massive}, we develop a BPCM to efficiently leverage channel correlation. Specifically, in the $(u,v)$-th sub-block, an affine function is used as an approximation of the channel coefficient:
\begin{gather} 
	h_{u,v,k,m,t,n} = c_{u,v,k,m} \gamma + d_{u,v,k,m} \lambda + q_{u,v,k,m} + \Delta_{u,v,k,m,t,n}, 
\end{gather}
where $c_{u,v,k,m}$, $d_{u,v,k,m}$ and $q_{u,v,k,m}$ represent the time-domain slope, the frequency-domain slope and the mean of the (u,v)-th planar sub-block, respectively; $\gamma$ and $\lambda$ are the bias of the time domain and frequency domain, respectively; $\Delta_{u,v,k,m,t,n}$ is the model mismatch error. 

Define time-domain biases vector $\mathbf e_1 = {\rm vec}((\bm \gamma_{1},..., \bm\gamma_{\frac{N}{V}})^T) $ $ \in \mathbb C^{\frac{NT}{VU} \times 1} $ with $ \bm\gamma_{i} = [-\frac{T}{2U}+1,..., \frac{T}{2U}]^T  \in \mathbb C^{\frac{T}{U} \times 1}$, $i = 1,...,\frac{N}{V}$, where ${\rm vec(\cdot)}$ represents vectorizing a matrix column by column; define frequency-domain biases vector $\mathbf e_2 = {\rm vec}(\bm \lambda_1 ,...,\bm \lambda_\frac{T}{U} ) \in \mathbb C^{\frac{NT}{VU} \times 1} $ with $ \bm \lambda_{j} = [-\frac{N}{2V}+1, ...,\frac{N}{2V}]^T \in \mathbb C^{\frac{N}{V} \times 1}$, $j = 1,...,\frac{T}{U}$; define $\mathbf 1_\frac{NT}{VU}$ as an all-one vector of length $\frac{NT}{VU}$. The block-wise planar model of $\mathbf H_{u,v,k}$ is given by
\begin{equation} \label{el_H}
	\mathbf H_{u,v,k} = \mathbf e_1 \mathbf c_{u,v,k}^T + \mathbf e_2 \mathbf d_{u,v,k}^T +  \mathbf 1_\frac{NT}{VU} \mathbf q_{u,v,k}^T + \mathbf \Delta_{u,v,k},
\end{equation}
where $\mathbf c_{u,v,k} = [c_{u,v,k,1},..., c_{u,v,k,M}]^T$; $\mathbf d_{u,v,k} = [d_{u,v,k,1},...,$ $d_{u,v,k,M}]^T$; $\mathbf q_{u,v,k} = [q_{u,v,k,1},...,q_{u,v,k,M}]^T$; $\mathbf \Delta_{u,v,k} \in \mathbb C^{\frac{NT}{VU} \times M}$. Donote $\mathbf X_{u,v,k} = {\rm diag} (\mathbf x_{u,v,k})$. Substituting \eqref{el_H} into \eqref{SystemModelMatrix}, we obtain the block-wise planar system model as
\begin{align}
	\label{eq_model}
	\mathbf Y_{u,v} = \mathbf A_{u,v}  \mathbf C_{u,v} + \mathbf B_{u,v}  \mathbf D_{u,v} + \mathbf F_{u,v}  \mathbf Q_{u,v} + \mathbf Z_{u,v},
\end{align}
where $\mathbf A_{u,v} = [\mathbf X_{u,v,1} \mathbf e_1,...,\mathbf X_{u,v,K} \mathbf e_1]$; $\mathbf B_{u,v} = [\mathbf X_{u,v,1} \mathbf e_2,$ $...,  \mathbf X_{u,v,K} \mathbf e_2]$; $\mathbf F_{u,v} = [\mathbf X_{u,v,1} \mathbf 1_\frac{NT}{VU}, ..., \mathbf X_{u,v,K} \mathbf 1_\frac{NT}{VU}]$; $\mathbf C_{u,v} = [\mathbf 
c_{u,v,1} ,..., \mathbf c_{u,v,K} ]^T  \in \mathbb C^{ K \times M}$ is 
\textit{time-domain slope matrix};  $\mathbf D_{u,v} = [\mathbf 
d_{u,v,1},...,  \mathbf d_{u,v,K} ]^T  \in \mathbb C^{K \times M}$ is 
\textit{frequency-domain slope matrix};  $\mathbf Q_{u,v} = [\mathbf 
q_{u,v,1},..., \mathbf q_{u,v,K}]^T  \in \mathbb C^{K \times M}$ is \textit{mean matrix}; $\mathbf Z_{u,v} = \sum_{k=1}^{K} \mathbf X_{u,v,k} \mathbf \Delta_{u,v,k} + 
\mathbf W_{u,v}$ is the summation of the AWGN and the model mismatch error. From the central limit theorem, $\mathbf Z_{u,v}$ can be modeled as an AWGN matrix. 

\begin{figure}[h]
	\centering
	\includegraphics[width=3.0in]{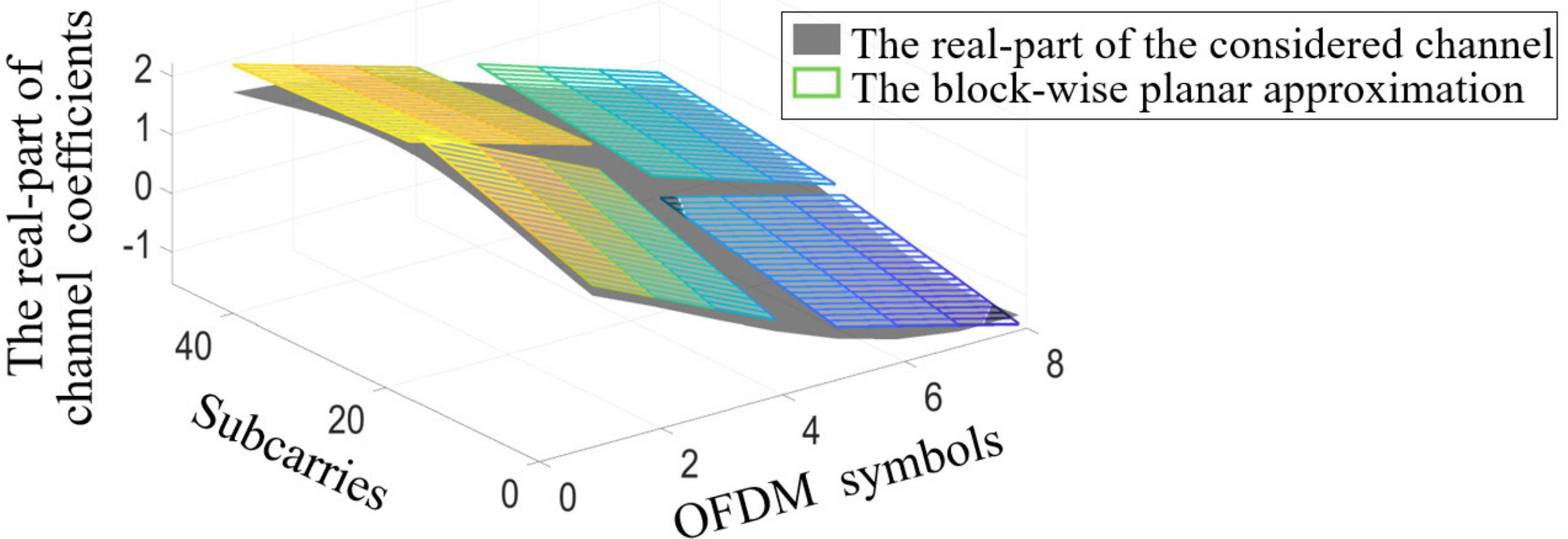}
	\caption{An example for the block-wise planar approximation of the real-part of the time-frequency-domain channel coefficients. The considered channel is a random realization drawn from the CDL-B channel model \cite{3GPP_TS_38_901_v17.0.0}.}
	\label{block}
\end{figure}

To facilitate the understanding of the BPCM, we show the time-frequency-domain channel coefficients in Fig. \ref{block}, where 48 subcarriers and 8 OFDM symbols are divided into 4 blocks with $U=2,V=2$. We see that the channel is accurately approximated by the BPCM.

Define $\mathbf A_{u,v}^{\mathrm P} = \mathbf S_{u,v} \mathbf A_{u,v} \in \mathbb C^{\frac{NT_{\mathrm P}}{VU} \times K} $, $\mathbf B_{u,v}^{\mathrm P} = \mathbf S_{u,v} \mathbf B_{u,v} \in \mathbb C^{\frac{NT_{\mathrm P}}{VU} \times K} $, $\mathbf F_{u,v}^{\mathrm P} = \mathbf S_{u,v} \mathbf F_{u,v} \in \mathbb C^{\frac{NT_{\mathrm P}}{VU} \times K} $, and $\mathbf Z_{u,v}^{\mathrm P} = \mathbf S_{u,v} \mathbf Z_{u,v} \in \mathbb C^{\frac{NT_{\mathrm P}}{VU} \times M}$. The received pilot signal is given by\footnote{To satisfy the solvability of the linear system, the number of observations within each sub-block should not be less than the number of variables to be estimated, i.e., $\frac{NT_p}{VU} \geq 3K$.}
\begin{align}
	\label{eq_model_pilot}
	\mathbf Y_{u,v}^{\mathrm P} = \mathbf A_{u,v}^{\mathrm P}  \mathbf C_{u,v} + \mathbf B_{u,v}^{\mathrm P}  \mathbf D_{u,v} + \mathbf F_{u,v}^{\mathrm P}  \mathbf Q_{u,v} + \mathbf Z_{u,v}^{\mathrm P}.
\end{align}

Based on \eqref{eq_model_pilot}, we use the LMMSE principle to estimate $\mathbf C_{u,v}$, $\mathbf D_{u,v}$ and $\mathbf Q_{u,v}$. The prior mean $\mathbf C_{u,v}^{\mathrm {Pri}}$, $\mathbf D_{u,v}^{\mathrm {Pri}}$, $\mathbf Q_{u,v}^{\mathrm {Pri}}$ are set to $\mathbf 0$, and the prior variance  $v_{\mathbf C_{u,v}}^{\mathrm {Pri}}$, $v_{\mathbf D_{u,v}}^{\mathrm {Pri}}$, $v_{\mathbf Q_{u,v}}^{\mathrm {Pri}}$ are assumed known and can be obtained from historical data. Then, the posterior mean $\mathbf C_{u,v}^{\mathrm {Post}}$ is obtained as
\begin{align}\label{cm}
	\mathbf C_{u,v}^{\mathrm {Post}} & =  \mathbf C_{u,v}^{\mathrm {Pri}}  + \  v_{\mathbf C_{u,v}}^{\mathrm {Pri}}  (\mathbf A_{u,v}^{\mathrm P})^H \mathbf {\boldsymbol \Sigma}_{u,v}^{-1} \times \notag \\ &  \left(\mathbf Y_{u,v}^{\mathrm P} - \mathbf A_{u,v}^{\mathrm P} \mathbf C_{u,v}^{\mathrm {Pri}} -\mathbf B_{u,v}^{\mathrm P} 	\mathbf D_{u,v}^{\mathrm {Pri}} -\mathbf F_{u,v}^{\mathrm P} 	\mathbf Q_{u,v}^{\mathrm {Pri}} \right),  
\end{align} 
where the covariance matrix $\mathbf {\boldsymbol \Sigma}_{u,v}$ is given by
\begin{align} \label{Conv}
	\mathbf {\boldsymbol \Sigma}_{u,v}  & = v_{\mathbf C_{u,v}}^{\mathrm {Pri}}\mathbf A_{u,v}^{\mathrm P} (\mathbf A_{u,v}^{\mathrm P})^H +  
	v_{\mathbf D_{u,v}}^{\mathrm {Pri}} \mathbf B_{u,v}^{\mathrm P} (\mathbf B_{u,v}^{\mathrm P})^H +  \notag \\ & ~~~~~~~~~~~~~~~~~~~~~~~~~
	v_{\mathbf Q_{u,v}}^{\mathrm {Pri}} \mathbf F_{u,v}^{\mathrm P} (\mathbf F_{u,v}^{\mathrm P})^H + \sigma_{\mathbf Z_{u,v}^{\mathrm P}}^2 \mathbf I.
\end{align}
Similarly to \eqref{cm}, we obtain $\mathbf D_{u,v}^{\mathrm {Post}}$ and $\mathbf Q_{u,v}^{\mathrm {Post}}$. Then, we reconstruct the pilot channel as
\begin{align} \label{Reconstruct}
	\hat {\mathbf {H}}_{u,v}^{\mathrm P} & = \rm diag(\mathbf S \mathbf e_1,...,\mathbf S \mathbf e_1) \mathbf C_{u,v}^{\mathrm {Post}} + \rm diag(\mathbf S \mathbf e_2,...,\mathbf S \mathbf e_2) \mathbf D_{u,v}^{\mathrm {Post}}   \notag \\ & ~~~~~~~~~~~~~~~ + \rm diag \left(\mathbf S \mathbf 1_{\frac{T_pN}{VU}},...,\mathbf S \mathbf 1_{\frac{T_pN}{VU}}\right) \mathbf Q_{u,v}^{\mathrm {Post}}. 
\end{align}

The pilot channels $\hat {\mathbf {H}}_{u,v}^{\mathrm P}, \forall u,v$, are rearranged to obtain the tensor $\hat {\bm {\mathcal H}}^{\mathrm P}_k \in \mathbb C^{ T_{\mathrm P} \times N \times M}, \forall k$ for subsequent filtering. Note that there are only three variables within each sub-block, and that the LMMSE principle is used for each sub-block separately. As such, the computational complexity is significantly reduced.

\subsection{Nonlinear Filtering Based on 3D-DRCN}
\begin{figure*}[h]
	\centering
	\includegraphics[width=6.1in]{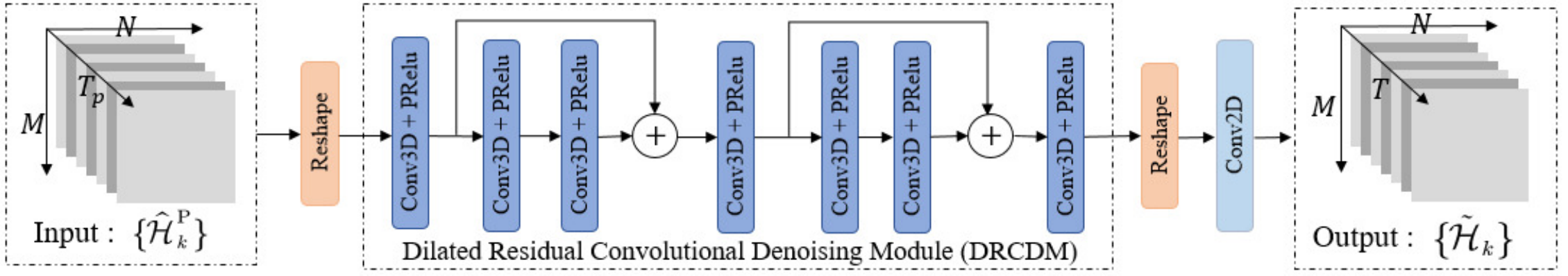}
	\caption{The block diagram of the proposed 3D-DRCN.  }
	\label{net}
\end{figure*}

We now describe the proposed 3D-DRCN, denoted by $\mathbf L\left(\cdot ;\boldsymbol{\theta} \right)$, that exploits the channel correlations in the time-frequency-space-domain in denoising and interpolation, where $\boldsymbol{\theta}$ represents the neural network parameters. Then the channel estimates for each user are obtained as
\begin{equation} \label{net_eq}
	\tilde{\boldsymbol{\mathcal H}}_k = \mathbf L\left(\hat {\bm {\mathcal {H}}}^{\mathrm P}_k ; \boldsymbol{\theta} \right)	,
\end{equation}
where $\tilde{\boldsymbol{\mathcal H}}_k$ are the estimates of the channel on both pilot and data symbols. The detailed structure of 3D-DRCN is shown in Fig. \ref{net}. After being reshaped into size $(1,T_{\mathrm P},N,M)$, the channel estimates from module A are sent into the dilated residual convolutional denoising module (DRCDM) for denoising, which is composed of seven 3D convolutional layers and seven parametric rectified linear unit (PRelu) activation functions. In DRCDM, the convolutional dilation rate is (1,4,4) and the size of the output feature map is $(T_{\mathrm P},N,M)$. The kernel size of the first to fourth 3D convolution layers and the fifth to seventh 3D convolution layers are $(7,7,5)$ and $(5,5,3)$, respectively. The number of output \textit{channels} of the first to third 3D convolution layers, the fourth to sixth 3D convolution layers, and the seventh 3D convolution layers are 1, 5, and 1, respectively.\footnote{Note that here "\textit{channel}" means the feature map in the convolutional operation of the neural network.} After denoising by the DRCDM, the channel estimates are reshaped into size $(T_{\mathrm P},N,M)$ for the last 2D convolution layer, which interpolates along the time domain and obtains the complete channel estimates $\tilde{\boldsymbol{\mathcal H}}_k$ with size $(T,N,M)$ on both pilot and data symbols. The kernel size of the last 2D convolution layer is $(5,3)$ and the dilation rate of the last 2D convolution layer is (4,4). Please note that compared with interpolation by filling zeros first and then denoising by the convolutional layer, interpolation by designing the number of \textit{channels} of the convolutional layer has the same result, but has lower computational complexity. The 3D-DRCN takes the L1 norm between the channel estimates and the true labels as the loss function for offline training. Adam is used as the optimizer and the learning rate is $10^{-4}$.

We note that the 3D-DRCN can be seen as an extension of the multi-scale parallel dilated convolutional neural network (MPDCNN) in \cite{LTMP}. The main differences are as follows. First, the 3D-DRCN uses the 3D convolution to exploit the time-frequency-space-domain correlations of the channel, while the MPDCNN in \cite{LTMP} is limited to the time-frequency domain. Second, the 3D-DRCN performs denoising and interpolation on the channel, while the MPDCNN in \cite{LTMP} is limited to denoising. Third, the 3D-DRCN is streamlined by removing redundant structures of the MPDCNN in \cite{LTMP}. Particularly, for 3D-DRCN, the space-domain kernel size is not larger than the time-frequency-domain kernel size, because the space-domain correlation of the channel is usually weaker than the time-frequency-domain correlations of the channel.

\subsection{Overall Algorithm}
The LBPCE performs the channel estimation with low complexity for time-varying frequency-selective MIMO-OFDM systems. We summarize LBPCE in Algorithm \ref{alg1}. In steps 1-3, we adopt the LMMSE principle to obtain the covariance matrix $\mathbf \Sigma_{u,v}$ and the estimates of the block-wise planar channel variables $\mathbf C_{u,v}$, $\mathbf D_{u,v}$, and $\mathbf Q_{u,v}$. In step 4, we reconstruct the pilot channel $\hat {\mathbf {H}}_{u,v}^{\mathrm P}, \forall u,v$ and rearrange it to obtain the tensor $\hat {\boldsymbol {\mathcal {H}}}_{k}^{\mathrm P}, \forall k$. In step 5, the 3D-DRCN filters  $\hat {\boldsymbol {\mathcal {H}}}_{k}^{\mathrm P}$ to output the  channel coefficients $\tilde{\boldsymbol{\mathcal H}}_k$ on both pilot and data symbols.

{ 
	\begin{algorithm}
		\small
		\caption{\label{alg1} LBPCE }
		\begin{algorithmic}
			
			\REQUIRE $\{\mathbf Y_{u,v}^{\mathrm P}\}$, $\{\mathbf x_{u,v,k}^{\mathrm P}\}$, $\{v_{\mathbf C_{u,v}}^{\mathrm {Pri}}\}$, $\{v_{\mathbf D_{u,v}}^{\mathrm {Pri}}\}$, $\{v_{\mathbf Q_{u,v}}^{\mathrm {Pri}}\}$.
					
			\textbf{\% Module A: Channel estimation with BPCM}
			
			\quad \ 1: Obtain the covariance matrix $\mathbf {\boldsymbol \Sigma}_{u,v}$ by \eqref{Conv}, $\forall u,v$.
			
			\quad \ 2: Obtain $\mathbf C_{u,v}^{\mathrm {Post}}$ by \eqref{cm}, $\forall u,v$.
			
		    \quad \ 3: Obtain $\mathbf D_{u,v}^{\mathrm {Post}}$ and $\mathbf Q_{u,v}^{\mathrm {Post}}$ similar to  \eqref{cm}, $\forall u,v$.
			
			\quad \ 4: Reconstruct $\hat {\mathbf {H}}_{u,v}^{\mathrm P}, \forall u,v$ by \eqref{Reconstruct}, and then rearrange it to obtain $\boldsymbol{\mathcal{\hat H}}^{\mathrm P}_{k}$, $\forall k$.
			
			\textbf{\% Module B: Nonlinear filtering based on 3D-DRCN }
			
			\quad \ 5: Obtain $ \tilde{\boldsymbol{\mathcal H}}_k$  by \eqref{net_eq}.
			
			\ENSURE The channel coefficients $ \tilde{\boldsymbol{\mathcal H}}_k$ on pilot and data symbols. 
		\end{algorithmic}
	\end{algorithm} 
}

We now analyze the computational complexity of LBPCE. We use the number of complex multiplications to measure the computational complexity and consider that 4 real-valued multiplications are equivalent to one complex multiplication. In module A, the channel estimation is performed in each sub-block separately, and the complex multiplications number of \eqref{cm}-\eqref{Reconstruct} is $ (\frac{NT_{\mathrm P}}{VU})^3 + 6K(\frac{NT_{\mathrm P}}{VU})^2 + 3KUVM(\frac{NT_{\mathrm P}}{VU})$. In module B, the 3D-DRCN filters the channel of each user separately. According to the detailed structure of 3D-DRCN shown in Fig. \ref{net}, the complex multiplications number is $(12170+30T)T_{\mathrm P}NMK/4$, which is linear to $T_{\mathrm P}$, $N$, $M$, and $K$. Thus, the number of multiplications of LBPCE is $(\frac{NT_{\mathrm P}}{VU})^3 + 6K(\frac{NT_{\mathrm P}}{VU})^2 + 3KUVM(\frac{NT_{\mathrm P}}{VU}) + (12170+30T)T_{\mathrm P}NMK/4$.

\section{Numerical Results}
In this section, we evaluate the performance of LBPCE. The simulation setup is as follows. The BS is equipped with $M = 64$ antennas and serves $K = 24$ users. $N=48$ OFDM subcarriers are allocated with subcarrier spacing $\Delta f = 30$ kHz. The number of OFDM symbols is $T=28$ with $T_{\mathrm P}=8$. For the network training, we use the CDL-B channel model \cite{3GPP_TS_38_901_v17.0.0} with 100-300 ns r.m.s. delay spread and consider the mobile speed of 80-120 km/h in the signal-to-noise ratio (SNR) range of 4-14 dB. For testing, we use the CDL-B channel with 129 ns r.m.s. delay spread and mobile speed of 100 km/h in the SNR range of 4-18 dB.

The baselines include CE-DNN\cite{ma2020data}, CE-CNN\cite{mashhadi2021pruning}, SF-CNN\cite{dong2019deep}, Model2\cite{melgar2022deep}, ICINet\cite{sun2021icinet}, LDAMP\cite{he2018deep}, LTMP \cite{LTMP}, 2DU\cite{marinberg2020study}, 3DFF\cite{marinberg2020study}, LS, 2$\times$1D LMMSE, 1D LMMSE, and Module A of LBPCE. Since works \cite{ma2020data} and \cite{mashhadi2021pruning} are a joint study of channel estimation and pilot design, which is beyond the research scope of our work, we only compare with their channel estimation sub-networks. Besides, we extend CE-DNN, CE-CNN, and SF-CNN for uplink multi-user scenarios and apply the LS method as a block-wise channel estimator before using these neural networks as denoisers. For the schemes that only design the pilot channel estimators, we employ linear interpolation to obtain the data channel estimates.

\begin{figure}[h]
	\centering
	\includegraphics[width=3.5in]{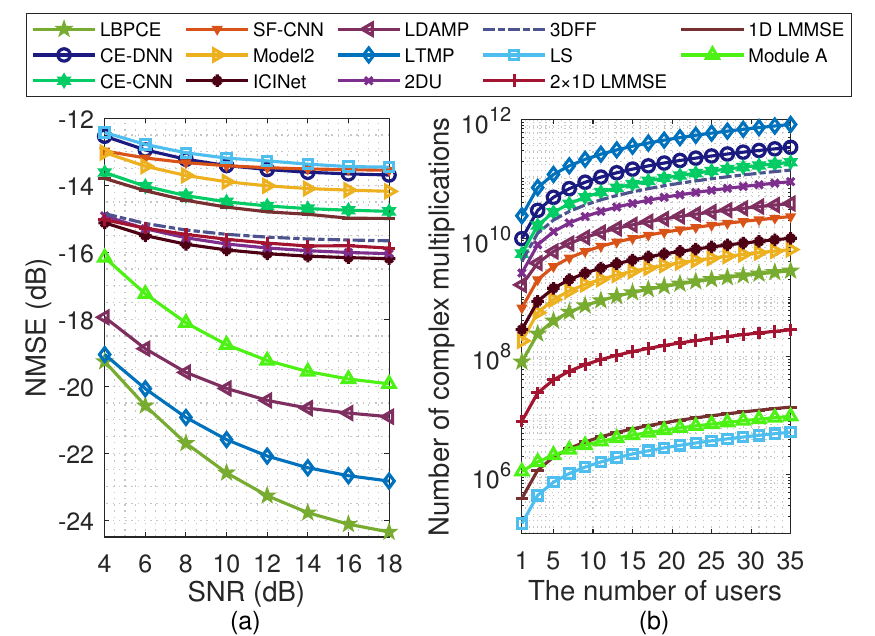}
	\caption{Channel estimation performance: (a) NMSE versus SNR; (b) the number of complex multiplications versus the number of users.}
	\label{CE1}
\end{figure}

For LBPCE, we set $U=2$ and $V=2$. Fig. \ref{CE1} shows the channel estimation MSE against SNR and the number of complex multiplications versus the number of users. We see that LBPCE significantly outperforms the CE-DNN, CE-CNN, SF-CNN, Model2, ICINet, 2DU, 3DFF, LS, 2$\times$1D LMMSE, and 1D LMMSE. Moreover, the NMSE floor of ICINet is -16 dB, while LBPCE reaches NMSE = -24 dB at SNR = 15 dB. Furthermore, LBPCE achieves better performance than LDAMP and LTMP while the computational complexity of LBPCE is reduced by more than one order of magnitude compared with LDAMP and by nearly three orders of magnitude compared with LTMP. Besides, the comparison with \enquote{Module A} verifies the refinement capability of module B.

To verify that 3D-DRCN achieves a favorable balance between performance and computational complexity, we carry out the ablation experiment as follows: (a) the 3D-DRCN-deep scheme that concatenates two DRCDMs; (b) the 3D-DRCN-parallel scheme that parallelizes three DRCDMs with convolutional dilation rates (1,1,1), (1,2,2), and (1,4,4); (c) the 3D-DRCN-\textit{channel} scheme that increases the input \textit{channel} numbers in the 2nd-5th and 6th-8th layers to 5 and 25, respectively; (d) the 3D-DRCN-dilation scheme that changes the dilation rates of all operations in the 3D-DRCN to 1. 

\begin{figure}[h]
	\centering
	\includegraphics[width=3.45in]{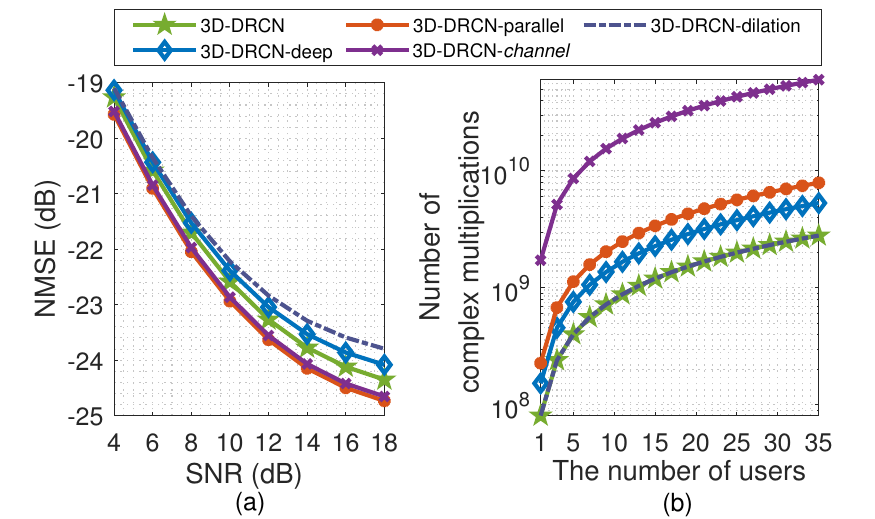}
	\caption{The ablation experiment performance of 3D-DRCN: (a) NMSE versus SNR; (b) the number of complex multiplications versus the number of users.}
	\label{CE2}
\end{figure}

The ablation experiment results are shown in Fig. \ref{CE2}. Compared with the 3D-DRCN-deep, the 3D-DRCN achieves better performance with a $50\%$ reduction in computational complexity, which indicates that the structure of 3D-DRCN-deep is too complex, resulting in performance loss due to overfitting. The 3D-DRCN-parallel and the 3D-DRCN-\textit{channel} have a slight NMSE gain at the cost of 3 times and 20 times increase in the computational complexity. Furthermore, compared with the 3D-DRCN-dilation, the 3D-DRCN improves performance without increasing the computational complexity. The above observations indicate that the 3D-DRCN strikes a favorable balance between performance and computational complexity.

\section{Conclusion}
In this paper, we proposed a low-complexity LBPCE for time-varying frequency-selective MIMO-OFDM systems. First, we established a BPCM which greatly reduced the number of variables to be determined in channel estimation. Second, we designed a low-complexity 3D-DRCN for channel filtering which explored the time-frequency-space-domain correlations of the channel. Simulation results demonstrate that LBPCE outperforms the state-of-the-art estimators and maintains low complexity.

\bibliographystyle{IEEEtran}
\bibliography{W0741}

\vfill
	
\end{document}